# Strong electron-hole symmetric Rashba spin-orbit coupling in graphene/monolayer transition metal dichalcogenide heterostructures

Bowen Yang[1], Mark Lohmann[1], David Barroso[2], Ingrid Liao[2], Zhisheng Lin[1], Yawen Liu[1], Ludwig Bartels[2], Kenji Watanabe[3], Takashi Taniguchi[3] & Jing Shi[1]

[1]Department of Physics and Astronomy, University of California, Riverside, CA 92521

[2]Department of Chemistry and Materials Science & Engineering Program, University of California, Riverside, CA 92521

[3]National Institute for Materials Science, 1-1 Namiki, Tsukuba, 305-0044 Japan

Despite its extremely weak intrinsic spin-orbit coupling (SOC), graphene has been shown to acquire considerable SOC by proximity coupling with exfoliated transition metal dichalcogenides (TMDs). Here we demonstrate strong induced Rashba SOC in graphene that is proximity coupled to a monolayer TMD film, $MoS_2$ or $WSe_2$, grown by chemical vapor deposition with drastically different Fermi level positions. Graphene/TMD heterostructures are fabricated with a pickup-transfer technique utilizing hexagonal boron nitride, which serves as a flat template to promote intimate contact and therefore a strong interfacial interaction between TMD and graphene as evidenced by quenching of the TMD photoluminescence. We observe strong induced graphene SOC that manifests itself in a pronounced weak anti-localization (WAL) effect in the graphene magnetoconductance. The spin relaxation rate extracted from the WAL analysis varies linearly with the momentum scattering time and is independent of the carrier type. This indicates a dominantly Dyakonov-Perel spin relaxation mechanism caused by the induced Rashba SOC. Our analysis yields a Rashba SOC energy of ~1.5 meV in graphene/$WSe_2$ and ~0.9 meV in graphene/$MoS_2$, respectively. The nearly electron-hole symmetric nature of the induced Rashba SOC provides a clue to possible underlying SOC mechanisms.

Since successful isolation of monolayer graphene [1], a wide variety of two-dimensional (2D) atomically layered materials have been investigated. These materials form a complete family ranging from insulators like hexagonal boron nitride (*h*-BN), semiconductors ($MoS_2$, $WSe_2$, etc.), semimetals ($WTe_2$, $MoTe_2$, TaAs [2-4], etc.), to superconductors ($NiSe_2$ [5]). In the 2D materials family, graphene stands out for its extraordinarily high mobility and ultra-small spin-orbit coupling (SOC) offering efficient transport of both electron charges and spins. However, the gapless band of graphene hinders applications such as transistors; the negligible SOC prevents novel quantum states from emerging at practically accessible temperatures such as the quantum spin Hall insulator [6], the first theoretically predicted topological insulator, and the quantum anomalous Hall state [7]. On the other hand, owing to the large band gaps (1.5 – 2.0 eV [8]), strong intrinsic SOC and spin-valley coupling [9, 10], and the valley Hall effect [11], transition metal dichalcogenides (TMDs) such as $MoS_2$ and $WSe_2$ generate research interest for potential opto-electronic device applications. However, TMDs have relatively low mobility and high contact resistance which are unfavorable for all-electrical device applications, especially those based on quantum transport properties.

A natural way to take advantage of the complementary attributes of graphene and TMDs is to fabricate van der Waals heterostructures incorporating both. For example, when graphene is placed on $WS_2$, it acquires SOC via proximity coupling that leads to a weak-antilocalization (WAL) effect absent in standalone graphene [12-14]. In our previous work on graphene/$WS_2$ [12], we showed that a sub-linear relation holds between spin relaxation rate and momentum scattering time at relatively high carrier densities ($5 \times 10^{12}\ cm^{-2}$ on the hole side), suggesting that the Rashba SOC dominates the spin relaxation via the Dyakonov-Perel (DP) mechanism. We also estimated the Rashba SOC strength to be approximately 0.5 meV, which is at least one order of magnitude greater than the strength of the intrinsic SOC of graphene [15]. To realize the quantum anomalous Hall effect at high temperatures, even larger Rashba SOC is needed in addition to the exchange interaction [16]. Therefore, other TMD materials with larger atomic SOC such as $WSe_2$ are preferred [17]. In addition, by tuning the Fermi level towards the conduction band of TMD, the hybridization between graphene's $\pi$-band and TMD's $d$-band becomes stronger, and the SOC is expected to be stronger based on the current understanding [18]. Furthermore, several recent studies [19, 20] suggest that the interaction between graphene and $MoS_2$ is greatly enhanced when $MoS_2$ is "turned on", which can also lead to an enhancement of SOC in graphene. In this work, we choose chemical vapor deposition (CVD) grown monolayer hole-doped WSe$_2$ and electron-doped MoS$_2$ to examine these effects. In graphene/MoS$_2$, MoS$_2$ can indeed be "turned on" above a certain positive gate voltage.

In general, monolayer TMDs have better gate tunability as well as larger on-off ratio than multilayer TMD due to smaller density-of-states [21]. However, unlike graphene, monolayer TMD sheets have a low yield in isolation by mechanical exfoliation, and the resulting small flakes also make the alignment with graphene target flakes difficult. On the other hand, CVD grown TMDs do not have these shortcomings. Moreover, monolayer TMD films can be intentionally doped [22], and easily scaled up to centimeter size, which not only drastically simplifies the heterostructure fabrication process, but also suits better for device applications. Several techniques [23, 24] have been developed in previous studies which used polymers to transfer TMD sheets to various insulating substrates. These techniques worked well for picking up centimeter-scale TMD films, but the polymer films in touch with graphene are possible contamination sources due to baking and dissolving which is needed during transfer. To avoid using these polymers, we replace them with $h$-BN for the first time to directly pick up continuous multi-grain TMD films by leveraging the van der Waals interaction (more details in Supplementary Material). The $h$-BN flake used for this pickup-transfer technique can not only ensure a high yield (80%) of heterostructure fabrication, but also serve both as an encapsulating layer to prevent the TMD from degrading in ambient condition and as a robust dielectric medium for top gating. Therefore, with this dry transfer technique, the TMD films are protected from any exposure to polymers or solvent solutions. Additionally, the accurate stamping process gives us flexibility of identifying and picking up defect-free areas under an optical microscope, which results in better cohesion of TMD films with graphene.

Fig. 1(a) shows an optical image of a graphene/WSe$_2$/$h$-BN heterostructure. The white dashed line highlights the boundary between a graphene/WSe$_2$/$h$-BN stack on the left and a graphene-free WSe$_2$/$h$-BN stack on the right. Also visible is a gold electrode used for top gating. The Raman spectra shown in Fig. 1(b) and its inset clearly reveal the characteristic modes of WSe$_2$, $h$-BN, and graphene respectively. The absence of the $\sim 308\ cm^{-1}$ peak indicates that the WSe$_2$ film is a monolayer [25]. Since monolayer WSe$_2$ on a dielectric has a direct band gap [8] which gives rise to a strong photoluminescence (PL) response, PL mapping can be used to locate the monolayer. The red region in Fig. 1(c) maps out the area with a strong PL peak located at 1.65 eV. It coincides with the monolayer WSe$_2$ area absent of graphene underneath. In the blue region where WSe$_2$ is in contact with graphene, the PL intensity is greatly suppressed (by a factor of ~ 20). It is known that the PL quenching occurs if graphene Dirac bands are aligned with the band gap of WSe$_2$ [26] (as shown in Fig. 1(d) inset) and electrons can freely move between the two layers. Therefore, the quenched PL is indicative of strong proximity coupling between graphene and WSe$_2$. We have also examined the topography of the heterostructures with atomic force microscopy (AFM) (see Supplementary Material Fig. S1) and found bubbles (with a height of ~ 10 nm) formed between graphene and WSe$_2$/h-BN, as revealed by the lighter spots on the left side of Fig. 1(a) and the corresponding red dots in Fig. 1(c). This indicates that under these bubbles graphene is detached from WSe$_2$ so that the PL is restored. The presence of the bubbles reduces the overall proximity coupling which could result in an underestimation of the SOC strength extracted from WAL as will be discussed later.

The heterostructure is then patterned into Hall bars with standard e-beam lithography and inductively coupled plasma etching. Cr/Au is deposited by e-beam evaporation to the contact areas containing the edges of the etched Hall bars to form one-dimensional edge contacts [27]. The mobility of completed devices ranges from 7,000 to 12,000 $cm^2 V^{-1} s^{-1}$, limited by the Coulomb scattering [28] from SiO$_2$ substrate. Fig. 2(a) and 2(b) show the top gate voltage dependence of the conductance $G$ and the carrier density $n$ (inset) in SiO$_2$/graphene/WSe$_2$ and SiO$_2$/graphene/MoS$_2$ respectively. In Fig. 2(b) and its inset, the carrier density reaches the saturation value of $3 \times 10^{12}$ $cm^{-2}$ as the top gate voltage approaches ~ 8 V. At this gate voltage, the conductance starts to deviate from the linear trend. The saturation occurs on the positive gate voltage side because similar to exfoliated MoS$_2$ monolayer CVD MoS$_2$ is naturally electron-doped [29] (also see Supplementary Material Fig. S2) and the Fermi level in graphene/MoS$_2$ is close to the conduction band minimum of MoS$_2$. Due to the strong coupling, electrons loaded to graphene escape to MoS$_2$ which results in the graphene electron density saturation. In contrast, the CVD WSe$_2$ is naturally hole-doped [30]. Within the applied gate voltage range there is no sign of carrier density saturation, indicating that the Fermi level of graphene/WSe$_2$ is above the valence band maximum of WSe$_2$. As recently reported, the conductance state of MoS$_2$ has a significant effect on spin transport in graphene channel [19, 20]. The widely different Fermi level position in these two TMDs allows us to directly compare this effect on acquired SOC.

Magnetoconductance (MC) measurements are carried out using a closed-cycle refrigerator system with temperature down to 4 K. Due to the unique Berry phase of $\pi$ of the graphene pseudo-spin, in the absence of other significant decoherence processes, intravalley coherence alone should in principle give rise to negative MC, i.e., the WAL effect [31]. However, due to relatively strong intravalley decoherence and intervalley scattering, weak localization (WL) is usually observed at low temperatures in standalone graphene devices [32, 33]. In contrast, if a strong Rashba SOC is introduced, graphene acquires an additional $\pi$ phase in the wave-function, allowing WAL to emerge [12-14, 34]. Therefore, the WAL feature serves as a direct indicator of induced Rashba SOC.

Fig. 3(a) and 3(b) show the MC data observed at several representative gate voltages/carrier densities in graphene/WSe$_2$ (hole side) and graphene/MoS$_2$ (electron side) respectively. The universal conductance fluctuation (UCF) in these devices is strongly suppressed owing to the large size of the devices made of large-area CVD TMD films. Following reference [32], we further reduce the effect of UCF by averaging the MC curves taken within a narrow moving window of carrier density $\sim 1.5 \times 10^{11}\ cm^{-2}$. Additionally, the MC curves are symmetrized with respect to zero magnetic field to eliminate any mixed antisymmetric (e.g., Hall) signals. The presence of low-field peaks (negative MC) clearly reveals the WAL effect, indicating that graphene in both devices acquires significant Rashba SOC from the monolayer TMD. As the carrier density approaches zero, the dephasing rate increases because of enhanced electron-electron interaction [33]. Consequently, the WAL peaks simultaneously weaken and broaden before vanishing when the dephasing rate exceeds the spin relaxation rate near the Dirac point.

Fig. 3(c) and 3(d) show the systematic gate voltage dependence of the MC data observed in graphene/WSe$_2$ and graphene/MoS$_2$, respectively. The ridges in the middle of the two-dimensional plot (vertical white line around zero magnetic field) represent the WAL peaks shown in Fig. 3(a) and 3(b), and the red regions indicate the WL effect that was also observed in pristine graphene [32]. In these devices, the carrier density threshold below which the WAL disappears is significantly lowered by a factor of five compared to our previous work, i.e. from $1.5 \times 10^{12}\ cm^{-2}$ to $3 \times 10^{11}\ cm^{-2}$, suggesting a larger spin relaxation rate relative to the dephasing rate in the present device.

Previous studies reported either absent spin Hall effect [35] or no systematic WAL data [13] on the hole side. Here, no matter whether the carrier density saturates on the electron side or not, a common feature in both Fig. 3(c) and 3(d) is that the WAL effect is nearly electron-hole symmetric. Note that in the graphene/MoS$_2$ device the Dirac point of graphene is in the vicinity of the conduction band minimum or the defect states [35] of MoS$_2$. As a result, the band hybridization on the electron side is expected to be enhanced. This effect would result in highly asymmetric density-of-states (i.e. electron density saturation) as well as strong SOC, i.e. stronger WAL[18]. Although this picture is consistent with the observed electron density saturation in graphene/MoS$_2$, it fails to explain the absence of any clear WAL enhancement over the same gate voltage range. Hence, the electron-hole symmetric nature of the observed WAL defies expectations based on earlier works and calls for a better understanding of the phenomenon.

Now we quantitatively analyze the MC results. The spin relaxation rate $\tau_{SOC}^{-1}$ can be extracted by fitting the WAL data with the diamgrammtic perturbation theory developed for the diffusive transport regime [34]. In our devices, the mean-free-path is less than 0.2 μm and the theory is perfectly applicable. Here in monolayer TMD, spins are oriented perpendicular to the layer. Due to strong coupling between graphene and monolayer TMD, spins in graphene also adopt the same orientation. Therefore, two spin relaxation processes occur to the perpendicular spins, i.e., precession due to the in-plane Rashba field via the Dyakonov-Perel (DP) mechanism and spin-flip due to the out-of-plane Kane-Mele field via the Elliott-Yafet (EY) mechanism. Hence, the spin relaxation rates $\tau_R^{-1}$ and $\tau_{KM}^{-1}$ due to Rashba SOC and KM SOC are related to the momentum scattering rate $\tau_p^{-1}$ by $\tau_R^{-1} = 2\frac{\lambda_R^2}{\hbar^2}\tau_p$ [36] and $\tau_{KM}^{-1} = \frac{\lambda_I^2}{\epsilon_f^2}\tau_p^{-1}$ [37, 38], respectively, where λ$_R$ and λ$_I$ are the strength of Rashba SOC and Kane-Mele SOC, respectively. The additional valley-Zeeman coupling term, as was discussed in our previous work [12], does not relax the spin orientation by momentum scattering, thus the total spin relaxation rate $\tau_{SOC}^{-1}$ is just a sum of $\tau_R^{-1}$ and $\tau_{KM}^{-1}$. The WAL fitting also allows us to obtain $\tau_\varphi^{-1}$.

Fig. 4(a) shows $\tau_{SOC}^{-1}$ and $\tau_\varphi^{-1}$ extracted from graphene/WSe$_2$ as a function of the carrier density. Below $n \sim 3 \times 10^{11}\ cm^{-2}$, $\tau_\varphi^{-1}$ exceeds $\tau_{SOC}^{-1}$ and the WAL feature is no longer observable as discussed earlier. In graphene, since $\tau_p \propto \sqrt{n}$, the DP and EY mechanism obey opposite density dependences, i.e., $\tau_R^{-1}$ ($\tau_{KM}^{-1}$) dominates at high (low) densities. Note that even at the lowest densities, $\tau_{SOC}^{-1}$ still follows the monotonic downward trend, suggesting that the dominant spin relaxation is governed by the DP mechanism. The slight asymmetry of $\tau_{SOC}^{-1}$ between hole and electron sides is due to the mobility difference: ~10,000 $cm^2V^{-1}s^{-1}$ for holes and ~6,500 $cm^2V^{-1}s^{-1}$ for electrons. Fig. 4(b) plots the $\tau_{SOC}^{-1}$ as a function of $\tau_p$ in graphene/WSe$_2$ and graphene/MoS$_2$ samples for both electron and hole sides. In contrast to reference [14] where $\tau_{SOC}^{-1}$ does not have explicit dependence on $\tau_p$, our result shows a clear linear relation between $\tau_{SOC}^{-1}$ and $\tau_p$ over a wider range of carrier densities, i.e., from $3 \times 10^{11}$ $cm^{-2}$ to $5 \times 10^{12}$ $cm^{-2}$ [12], unveiling dominance of the DP mechanism by the Rashba SOC over the entire range. More strikingly, the slopes of the linear fits are approximately the same for electrons and holes, reflecting near-symmetry in the SOC strength in both devices. It should be emphasized that the symmetry holds regardless of whether carrier density reaches saturation or not. This is in stark contrast to the results of previous studies [35], and therefore argues against the sulfur-vacancy related hybridization mechanism.

From the linear fits, we extract the Rashba SOC strength to be 1.4 – 1.6 meV in graphene/WSe$_2$, and 0.8 – 0.9 meV in graphene/MoS$_2$, increased by a factor of 2 – 3 from our previous study. Since molybdenum has weaker atomic SOC than tungsten, it is not surprising that the SOC in graphene/MoS$_2$ is weaker compared to graphene/WSe$_2$. The highly symmetric Rashba SOC strength suggests that the hybridized band structure plays a less important role, at least in graphene/MoS$_2$. It is worth pointing out that the Rashba SOC strength obtained here is still impacted by the presence of bubbles trapped between graphene and TMD which unavoidably dilutes the WAL response in the MC. Indeed, in a bubble-free graphene/WSe$_2$ device we have observed a stronger Rashba SOC (more details in Supplementary Material). Further improvement in device fabrication to reduce bubbles will certainly lead to an even greater Rashba SOC from entire strongly coupled graphene/TMD devices.

In summary, we have studied the WAL of graphene closer to the charge neutrality point when it is proximity coupled to CVD grown monolayers of WSe$_2$ or MoS$_2$. We have found that the spin relaxation is governed by the DP mechanism based on significant acquired Rashba SOC (0.8 – 1.5 meV). This SOC strength has strong electron-hole symmetry in both systems despite the widely different positions of their Fermi levels. The greatly enhanced Rashba SOC and the clear electron-hole symmetry are both important for better understanding the physical origin of the proximity induced SOC.

Acknowledgements:

Graphene/TMD heterostructure device fabrication, transport measurements, and data analysis were supported by the DOE BES award No. DE-FG02-07ER46351. Construction of the pickup-transfer microscope and device characterization were supported by NSF-ECCS under Awards No. 1202559 NSF-ECCS and No. 1610447. TMD material preparation was supported by C-SPIN, one of six centers supported by the STARnet phase of the Focus Center Research Program (FCRP), a Semiconductor Research Corporation program sponsored by MARCO and DARPA.

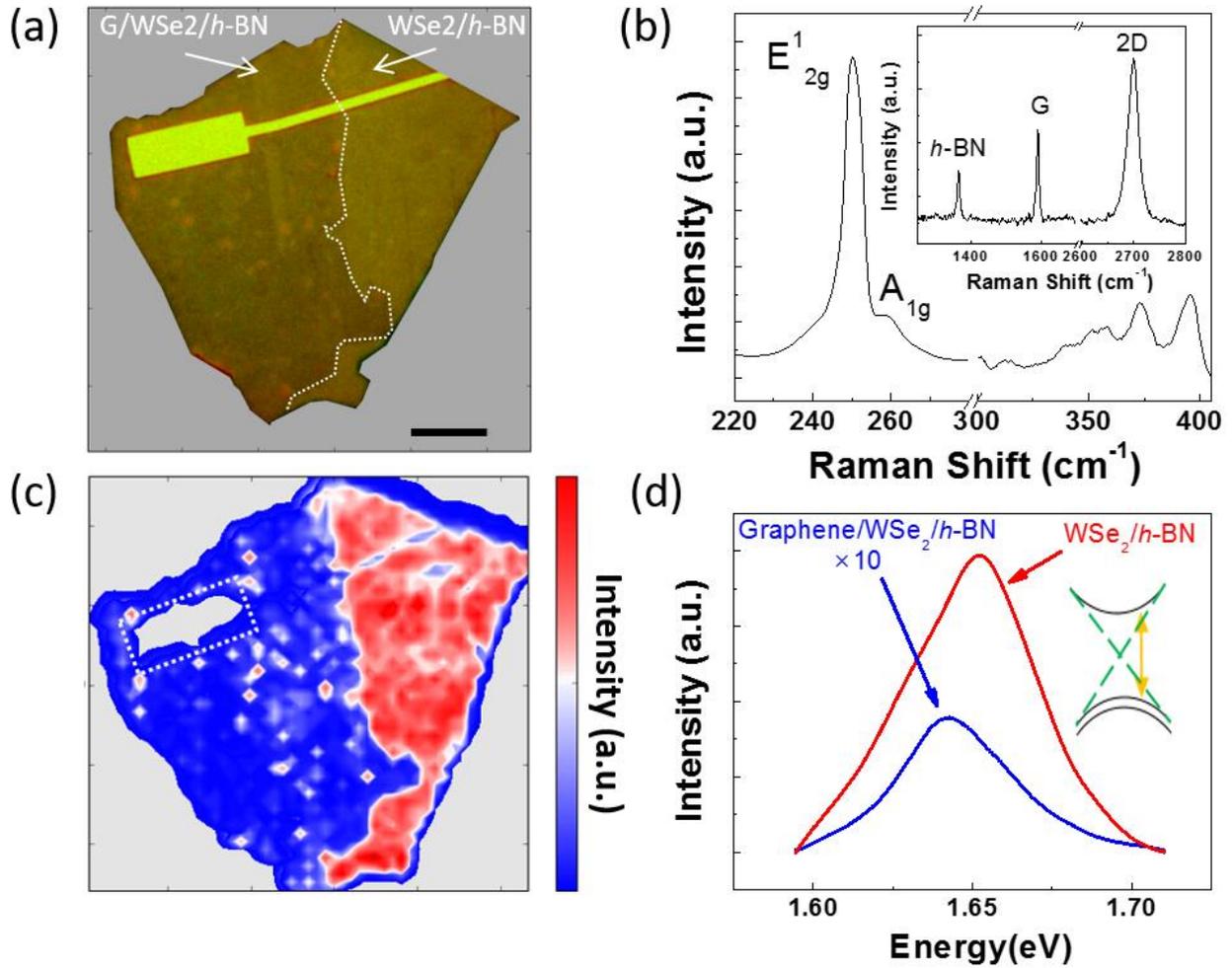

FIG. 1 (color online). (a) Optical image of graphene/$WSe_2$/$h$-BN (from bottom to top) heterostructure. The Si/$SiO_2$ background surrounding the heterostructure is removed for clarity. White dotted line delineates the boundary between areas with (left) and without graphene (right). The scale bar is 10 μm. (b) Raman spectra of the right area in (a). Inset: characteristic Raman peaks of $h$-BN and graphene. (c) PL mapping of the sample shown in (a). White dashed rectangle indicates the area covered by the gold pad in (a). (d) PL spectra taken in blue and red regions in (c). Note the PL intensity from the left with graphene underneath is magnified by 10 times for comparison. Inset: graphene Dirac bands present in the gap of $WSe_2$.

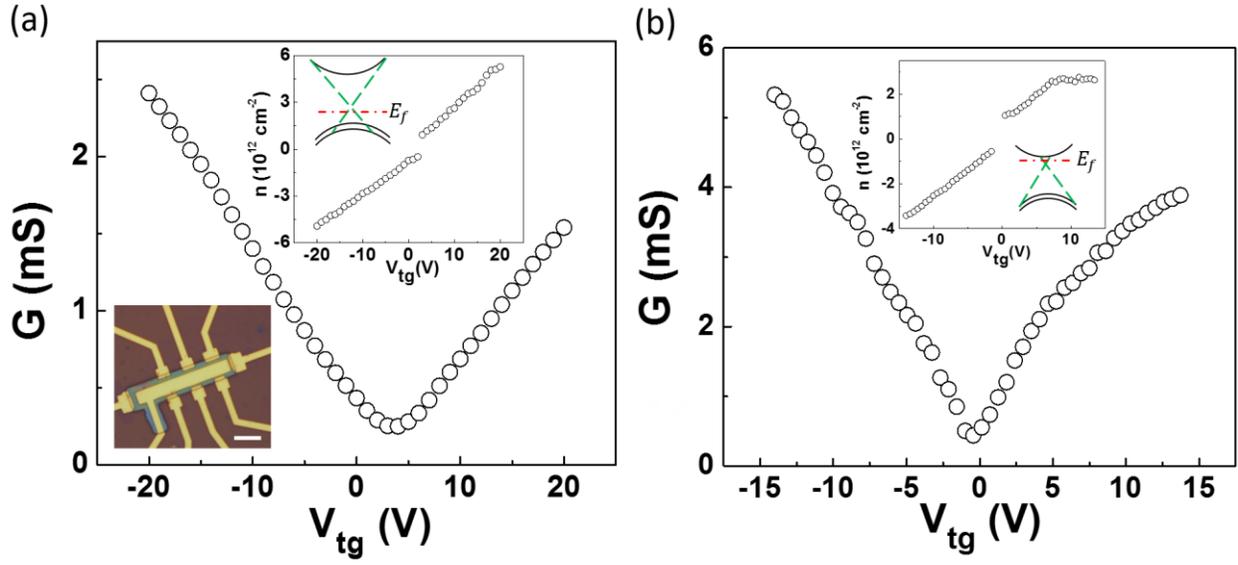

FIG. 2 (color online). Conductance $G$ and carrier density $n$ (inset) of graphene/$WSe_2$ sample 1 (a) and graphene/$MoS_2$ (b) as a function of top gate voltage. Bottom left inset in (a): optical image of the device. Scale bar is 10 $\mu m$. Insets in the carrier density plot draw the relative Fermi level positions in graphene/$WSe_2$ (a) and graphene/$MoS_2$ (b).

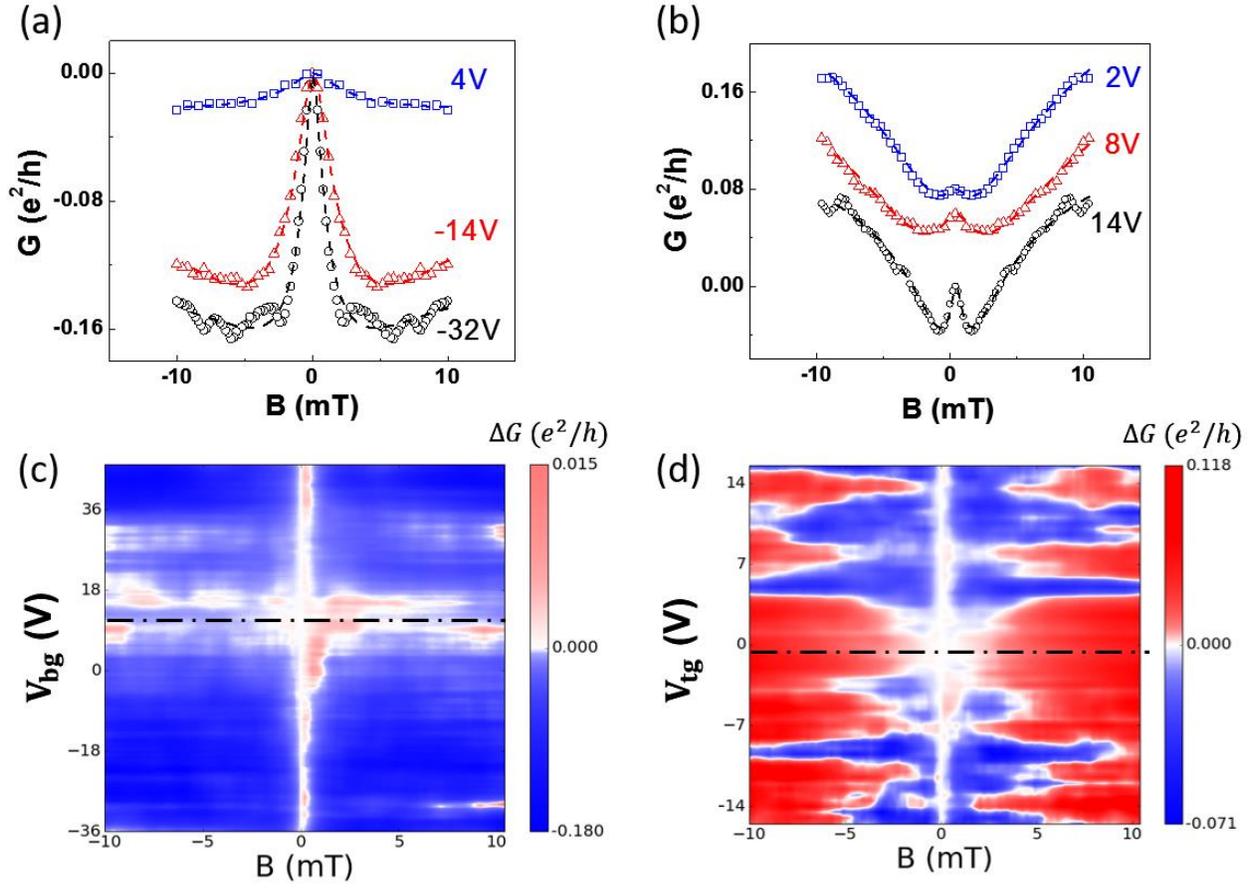

FIG. 3 (color online). MC curves, *G* vs. *B*, taken at different back gate voltages for graphene/WSe$_2$ sample 2 (a) and graphene/MoS$_2$ (b). MC curves in (b) are shifted vertically for clarity. Dash lines are the fits using equation (1). (c) and (d) are plots over a wide range of back gate voltages and magnetic fields from -10 to 10 mT corresponding to samples used in (a) and (b), respectively. Charge neutrality points are indicated by the black dotted dash lines.

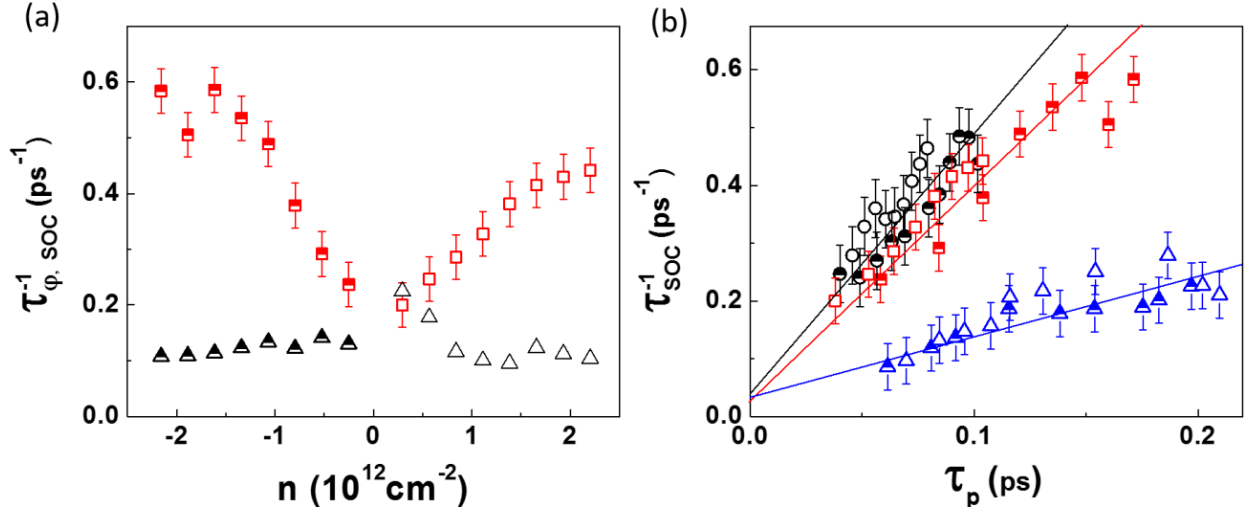

FIG. 4 (color online). (a) Spin relaxation rate $\tau_{SOC}^{-1}$ (squares) and dephasing rate $\tau_{\varphi}^{-1}$ (triangles) as a function of carrier density $n$. These rates are extracted from graphene/WSe$_2$ sample 2. Half filled (open) symbols stand for hole (electron) side. (b) $\tau_{SOC}^{-1}$ as a function of $\tau_p$ in three different samples: circles for graphene/WSe$_2$ sample 1, squares for graphene/WSe$_2$ sample 2, and triangles for graphene/MoS$_2$. Open and half-filled symbols denote the electron and hole side, respectively. Solid lines are guidelines to the eye.

References

[1] K. S. Novoselov *et al.*, Science **306**, 666 (2004).

[2] T. A. Empante *et al.*, ACS Nano, **11**, 900 (2017).

[3] B. Q. Lv, *et al.*, Phys. Rev. X **5**, 031013 (2015).

[4] S. Y. Xu *et al.*, Science **347**, 294 (2015).

[5] Y. Cao, *et al.*, Nano Lett. **15**, 4914 (2015).

[6] C. L. Kane and E. J. Mele, Phys. Rev. Lett. **95**, 226801 (2005).

[7] Z. H. Qiao, Phys. Rev. B **82**, 161414(R) (2010).

[8] Z. Y. Zhu, Y. C. Cheng, and U. Schwingenschlögl, Phys. Rev. B **84**, 153402 (2011).

[9] D. Xiao, W. Yao, and Q. Niu, Phys. Rev. Lett. **99**, 236809 (2007).

[10] D. Xiao, G. Liu, W. Feng, X. Xu, and W. Yao, Phys. Rev. Lett. **108**, 196802 (2012).

[11] K. F. Mak, K. L. Mcgill, J. Park, P. L. Mceuen, Science **344**, 1489 (2014).

[12] B. W. Yang *et al.*, 2D Mater. **3**, 031012 (2016).

[13] Z. Wang *et al.*, Nat. Commun. **6**, 8339 (2015).

[14] Z. Wang *et al.*, Phys. Rev. X **6**, 041020 (2016).

[15] H. Min *et al.*, Rev. B **74**, 165310 (2006).

[16] Z. Wang, C. Tang, R. Sachs, Y. Barlas, and J. Shi, Phys. Rev. Lett. **114**, 016603 (2015).

[17] M. Gmitra, D. Kochan, P. Högl, and J. Fabian, Phys. Rev. B **93**, 155104 (2016).

[18] M. Gmitra and J. Fabian, Phys. Rev. B **92**, 155403 (2015).

[19] W. Yan, *et al.*, Nat. Commun. **7**, 13372 (2016).

[20] A. Dankert, and S. P. Dash, arXiv:1610.06326 (2016).

[21] B. Radisavljevic *et al.*, Nat. Nanotechnol. **6**, 147 (2011).

[22] J. Suh *et al.*, Nano Lett. **14**, 6976 (2014).

[23] A. L. Elı́as, *et al.*, ACS Nano **7**, 5235 (2013).

[24] A. Gurarslan, *et al.*, ACS Nano **8**, 11522 (2014).

[25] H. Li, *et al.*, Small **9**, 1974 (2013).

[26] C. J. Shih, ACS Nano **8**, 5790 (2014).

[27] L. Wang, Science **342**, 614 (2013).

[28] J.-H. Chen *et al.*, Nat. Phys. **4**, 377 (2008).

[29] E. Preciado *et al.*, Nat. Commun. **6**, 8593 (2015).

[30] B. L. Liu *et al.*, ACS Nano **9**, 6119 (2015).

[31] T. Ando and T. Nakanishi, J. Phys. Soc. Jpn. **67**, 1704 (1998).

[32] F. V. Tikhonenko *et al.*, Phys. Rev. Lett. **100**, 056802 (2008).

[33] F. V. Tikhonenko *et al.*, Phys. Rev. Lett. **103**, 226801 (2009).

[34] E. McCann and V. I. Fal'ko, Phys. Rev. Lett. **108**, 166606 (2012).

[35] A. Avsar *et al.*, Nat. Commun. **5**, 4875 (2014).

[36] M. I. Dyakonov and V. I. Perel, Sov. Phys. Solid State **13**, 3023 (1971).

[37] P. G. Elliott, Phys. Rev. **96**, 266 (1954).

[38] Y. Yafet, Solid State Physics (Academic, New York, 1963).